\documentclass[12pt,preprint]{aastex}

\shorttitle{Electron Acceleration in Sagittarius A*}

\begin{document}


\title{Electron Acceleration around the Supermassive Black Hole at the Galactic Center}


\author{Siming Liu\altaffilmark{1}, Vah\'{e} Petrosian\altaffilmark{2} and 
Fulvio Melia\altaffilmark{3}}


\altaffiltext{1}{Center for Space Science and Astrophysics, Department of Physics, Stanford
University, Stanford, CA 94305; email: liusm@stanford.edu}
\altaffiltext{2}{Department of Physics and Applied Physics, Stanford University, Stanford, 
CA 94305}
\altaffiltext{3}{Physics Department and Steward Observatory, The University of Arizona, 
Tucson, AZ 85721; Sir Thomas Lyle Fellow and Miegunyah Fellow.}


\begin{abstract}

The recent detection of variable infrared emission from Sagittarius A*, combined with its
previously observed flare activity in X-rays, provides compelling evidence that at least a
portion of this object's emission is produced by nonthermal electrons. We show here that
acceleration of electrons by plasma wave turbulence in hot gases near the black hole's
event horizon can account both for Sagittarius A*'s mm and shorter wavelengths emission in
the quiescent state, and for the infrared and X-ray flares, induced either via an
enhancement of the mass accretion rate onto the black hole or by a reorganization of the
magnetic field coupled to the accretion gas. The acceleration model proposed here produces
distinct flare spectra that may be compared with future coordinated multi-wavelength
observations. We further suggest that the diffusion of high energy electrons away from the
acceleration site toward larger radii might be able to account for the observed
characteristics of Sagittarius A*'s emission at cm and longer wavelengths.

\end{abstract}






\section{Introduction}

The observation of stellar motions within light-days of Sagittarius A*, a compact radio
source at the Galactic Center \citep{Balick74}, has provided compelling evidence that this
source is the radiative manifestation of a $\sim$ four million solar mass black hole
\citep{Schodel02, Ghez031}. The recently detected infrared emission and flare activity have
provided an additional evidence that this source is powered by a hot gas accreting onto the 
black hole \citep{Baganoff01, Goldwurm03, Baganoff032, Porquet03, Zhao04, Ghez03}. The 
quasi-periodic near-infrared variability may be an indication that the gas flared up before 
spiraling into the black hole \citep{Genzel03}. It is now generally agreed that the radio 
and infrared emission, and the flares, are likely produced by nonthermal high energy 
electrons (Liu \& Melia 2001; Genzel et al. 2003;  see also Mahadevan 1998).  However, the 
exact nature of the mechanism responsible for the acceleration of the electrons has not 
been addressed. This has given rise to diverse interpretations of the observations with 
assumed spectra of the accelerated electrons \citep{Markoff01, Liu02a, Nayakshin03, 
Yuan03}.

In this letter, we show that the mechanism producing high-energy particles in solar flares
works equally well in hot plasmas near the black hole. The solar flare model is based on a
second order Fermi acceleration process or a stochastic acceleration (SA) of particles by
interacting resonantly with plasma waves or turbulence (PWT) generated via an MHD
dissipation process (see e.g. Miller \& Ramaty 1987; Hamilton \& Petrosian 1992;  
Petrosian \& Liu 2004). In Sagittarius A*, nonthermal particles can be produced by the
turbulence expected to be induced by the magneto-rotational instability in the accretion
torus (Balbus \& Hawley 1991; Melia, Liu \& Coker 2001). The radiation emitted at the
acceleration site can explain the quiescent state mm and shorter wavelength observations.
Solar flares are energized by the process of magnetic reconnection during the dynamical
evolution of the coronal magnetic field. Similar processes in Sagittarius A* can release
energy in a small region and produce what we call a local event. A global flare can be
induced by an MHD fluctuation in the accretion torus or an enhancement of the accretion
rate onto the black hole. The emission spectra from these two energization mechanisms are
quite different, and can explain the distinct flare behaviors.

In \S\ \ref{SA}, we outline the theory of SA. Its application to Sagittarius A* is 
presented in \S\ \ref{sgra}. The main results of this letter are summarized in \S\ 
\ref{dis}, where we also discuss consequences of energetic electrons escaping the 
acceleration site. If such electrons can diffuse toward larger radii, they may account for 
Sagittarius A*'s emission at cm and longer wavelengths and has the potential to explain
the observed linear and circular polarization characteristics, and variabilities of this 
source.

\section{Stochastic Electron Acceleration}
\label{SA}

In the theory of SA \citep{Petrosian04}, particles are accelerated from a background plasma
to higher energies by interacting resonantly with PWT. The particle distribution and the
consequent emission spectrum are determined by the acceleration ($D_{\rm EE}$) and scattering
($D_{\mu\mu}$) rates due to the PWT, the energy loss rate ${\dot E}_{\rm L}$, and the spatial
diffusion time $T_{\rm esc}$.  These rates and time are determined by the magnetic field $B$,
density $n$ and particle distribution of the background plasma, the spectrum and intensity of
the turbulence, and the size of the acceleration site $R$.

The wave modes in a plasma are described by the dispersion relation $\omega=\omega({\bf k})$,
which depends primarily on the plasma parameter $\alpha = \omega_{\rm pe}/\Omega_{\rm e} =
(4\pi m_{\rm e}n)^{1/2}c/B\,,$ the ratio of the electron plasma frequency $\omega_{\rm pe} =
(4\pi n e^2/m_{\rm e})^{1/2}$ to the nonrelativistic electron gyrofrequency $\Omega_{\rm e} =
eB/m_{\rm e} c$, where $m_{\rm e}$, $c$ and $e$ are the electron mass, the speed of light,
and the elemental charge unit, respectively.  Electrons with Lorentz factor $\gamma$,
velocity $v$, and pitch angle cosine $\mu$ couple strongly with the wave satisfying the
resonance condition: $\omega=k_{||}v\mu+\gamma^{-1}\,,$ where $\omega$ and $k_{||}$ are the
wave frequency and wave vector parallel to the large scale magnetic field (only the first
harmonic of the gyrofrequency is considered here).  Electrons with higher energies resonate
with waves with smaller wave numbers ($k$) corresponding to larger spatial scales.  

In the following discussion, we adopt a turbulence spectrum characterized by a broken power 
law with a lower wavenumber cutoff at $k_{\rm min}$, corresponding to the scale where the 
turbulence is generated ($<R$), and a break wavenumber $k_{\rm max}$ above which the wave 
dissipation is fast. We will study the SA by a PWT propagating parallel to the large scale 
magnetic field.  The characteristic interaction time $\tau_{\rm p}$ is given by
$\tau_{\rm p}^{-1} = (\pi/2)\Omega_{\rm e}(8\pi{\cal E}_{\rm tot}/
B^2)(q-1)k_{\rm min}^{q-1}\,,$
where $q>1$ is the turbulence spectral index in the inertial range and ${\cal E}_{\rm
tot} $ is the total turbulence energy density.  Relativistic electrons interacting with 
waves in a plasma with $\alpha>(m_{\rm e}/m_{\rm p})^{1/2}\simeq 1/43$, where $m_{\rm p}$ 
is the proton mass, will have an isotropic distribution because of the short scattering 
time $\tau_{\rm sc}=1/D_{\mu\mu}$.  In such plasmas electrons can gain energy by interacting 
with the waves and lose energy either via Coulomb collisions with background plasma 
particles (at lower energies) or by radiative processes (at higher energies).  The electrons 
also diffuse spatially and leave the region on a characteristic time $T_{\rm esc}\sim 
R^2/\beta^2c^2\tau_{\rm sc}$.  The spatially integrated electron distribution $N(E)$, as a 
function of the kinetic energy $E$, satisfies the well known diffusion-convection equation:
\begin{equation}
{\partial N\over\partial t}= {\partial^2\over \partial E^2}(D_{EE} N) + {\partial\over
\partial E}[({\dot E}_{\rm L}-A) N] -{N\over T_{\rm esc}} + Q\,, \label{dceq}   
\end{equation}
where $D_{EE}$ and $A$ are obtained from the wave-particle interaction rates and $Q$ is a 
source term.  In Sagittarius A*, where the photon energy density is at least one order of 
magnitude lower than the magnetic field energy density, the radiative loss of relativistic 
electrons is dominated by synchrotron process, for which 
\begin{equation}
\tau_{\rm loss} = E/\dot{E}_{\rm L} = {\gamma-1\over 4\pi r_{\rm o}^2 c
n(\ln{\Lambda}/\beta + 4\beta^2\gamma^2/9\alpha^2)}\,, \label{loss}
\end{equation}
where $\ln{\Lambda} \simeq 20$ and the first and second term in the denominator give the 
Coulomb collision and synchrotron loss rate, respectively.

\section{Application to Sagittarius A*}
\label{sgra}

To determine the exact turbulence spectrum, one needs to solve the coupled kinetic equations
of the waves and particles and their solutions depend on the wave generation, cascade,
dissipation processes and the distribution of the background particles (see e.g. Miller,
LaRosa \& Moore 1996).  The lack of a well established theory for the MHD turbulence renders
such a treatment unpractical.  As an approximation, we set $k_{\rm min} = 2\pi\eta/R$ with 
$\eta>1$, where the characteristic variation length of the large scale magnetic field 
$R/\eta$ is related to the turbulence generation length scale.  To estimate $k_{\rm max}$, we 
note that the wave damping is efficient when the waves start to resonate with the background 
particles.  From the resonance condition, one has $k_{\rm max}\sim \Omega_{\rm 
e}/(c<\gamma>)$, where $<\gamma>$ is the mean Lorentz factor of the background electrons, 
which is expected to be $\sim 50$ near the black hole of Sagittarius A*.  For $k_{\rm min}\le 
k\le k_{\rm max}$ the turbulence most likely has a Kolmogorov spectrum with an index $q=5/3$ 
and above $k_{\rm max}$ the index changes to 4.0, a typical value obtained from MHD 
simulations \citep{Hawley95, Vestuto03}.  MHD simulations \citep{Hawley95, Hirose04} also 
suggest that the magnetic field energy density is about a few percent of the energy density 
of the background gas, which means that $\alpha\sim 1$ (note that $\alpha^2 = 0.5 nm_{\rm e}
c^2/[B^2/8\pi]$), and that ${\cal E}_{\rm tot} \lesssim B^2/8\pi$. The accretion flow in 
Sagittarius A* is collisionless; it is not clear how the particle distribution of the 
accretion plasma evolves from large to small radii, where it enters the acceleration site of 
strong dissipation and PWT.  We make the reasonable assumption that the source electrons have 
a relativistic Maxwellian distribution.  We therefore have ${\cal E}_{\rm tot} = a nk_{\rm 
b}T$ with $a\ll1$ as a model parameter.  The exact spatial diffusion time (or $T_{\rm esc}$) 
depends on the structure of the magnetic field \citep{Hirose04} and the scattering rate.  We 
set $T_{\rm esc} = \tau_{\rm tr}^2/\tau_{\rm sc}+\tau_{\rm tr}$, where the transit time 
$\tau_{\rm tr}=\eta R/v$ determines the escape time when the scattering time $\tau_{\rm
sc}\gg\tau_{\rm tr}$ \citep{Petrosian99, Petrosian04}. 

During big flares, the physical conditions are expected to undergo dramatic changes, which 
may affect the turbulence spectrum and the energy partition among the gas, turbulence, and 
magnetic field.  These processes are not well understood.  For the purpose of illustration, 
we will make the following reasonable assumption to reduce the number of parameters. We set 
$\eta = 2$, $k_{\rm max}=0.02\,\Omega_{\rm e}/c=1.2\times 10^{-5}(B/G)$ cm$^{-1}$, $\alpha=1$ 
and $a= 0.008$. This leaves only three independent parameters; here we choose $R$, $n$ and 
$\tau_{\rm p}$. All other quantities can be derived from them. 


To calculate the emission predicted by the model, one also needs to specify the geometry of
the source. We will assume that the source is uniform and spherically symmetric with a
radius $R$, which may or may not be centered on the black hole. (Note that although the
actual geometry is inferred to be Keplerian for a global acceleration region---see, e.g.,
Melia et al. 2001a---its differences from the simplified geometry adopted here are expected
to change the required model parameters only slightly. This is a fair assumption for local
fluctuations.) For flares induced by an enhanced accretion rate, the density $n$ can
increase dramatically while $R$ and $\tau_{\rm p}$ may not change much at all. For local
MHD fluctuations, all three parameters can experience significant changes \citep{Hawley95}.

\begin{figure}
\plotone{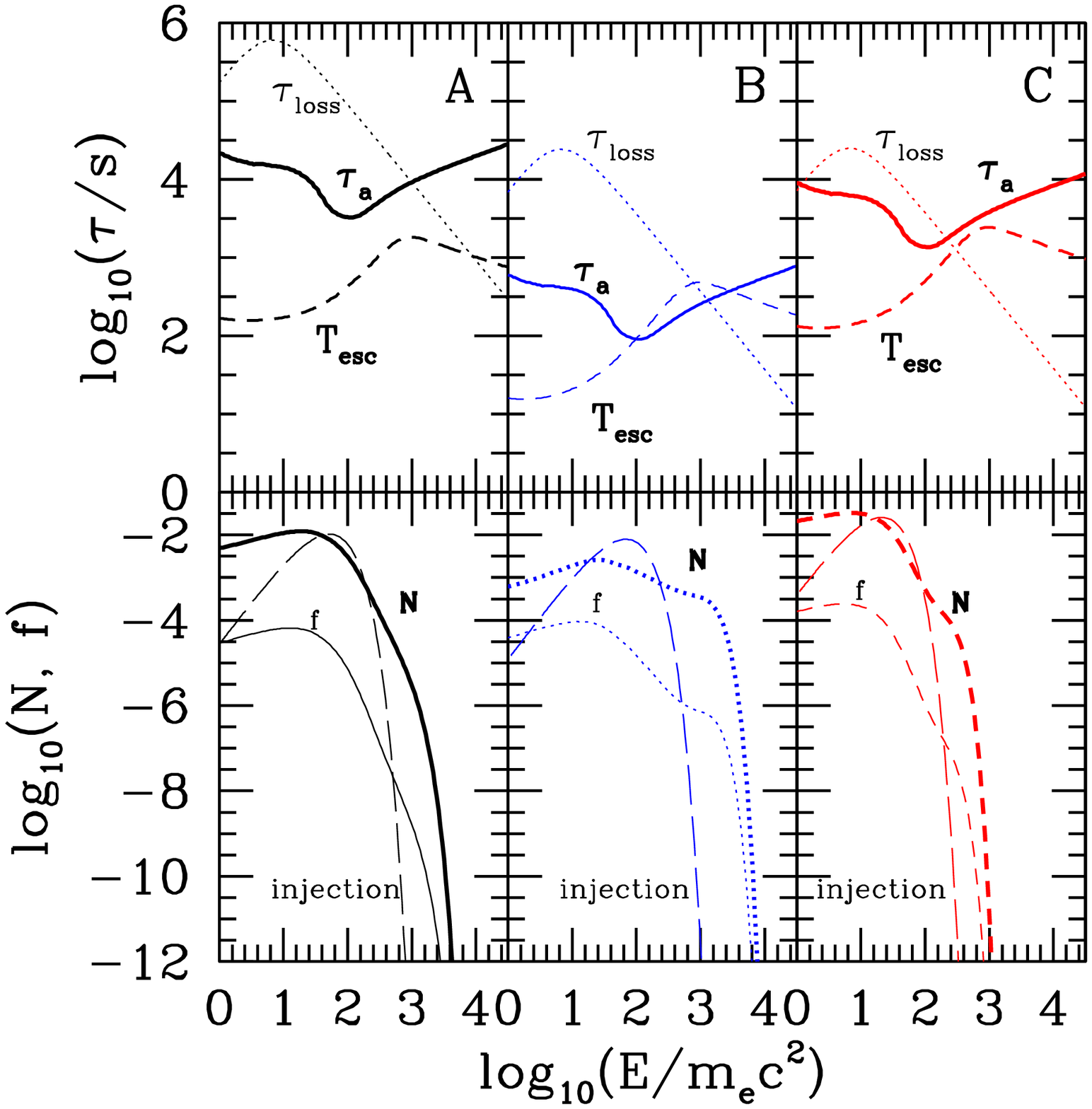}
\caption{
{\bf Top Panels:} The time scale for acceleration $\tau_{\rm a}=E/A$ (solid lines), escape 
(dashed lines) and loss $\tau_{\rm loss} = E/\dot{E}_{\rm L}$ 
(dotted lines) for Models A (left panel), B (middle panel) and C (right panel). The model 
parameters are given in Table \ref{tab:mods}. The loss term is dominated by 
Coulomb collisions below the peak and by synchrotron losses above 
it. {\bf Bottom Panels:} The normalized electron distributions N (thick lines) at the 
acceleration site and the corresponding escape fluxes $f = N/T_{\rm esc}$ (thin lines). 
The long dashed lines show the source Maxwellian electron distribution. See text for 
details.
\label{fig:acc}}
\end{figure}

\begin{figure}
\plotone{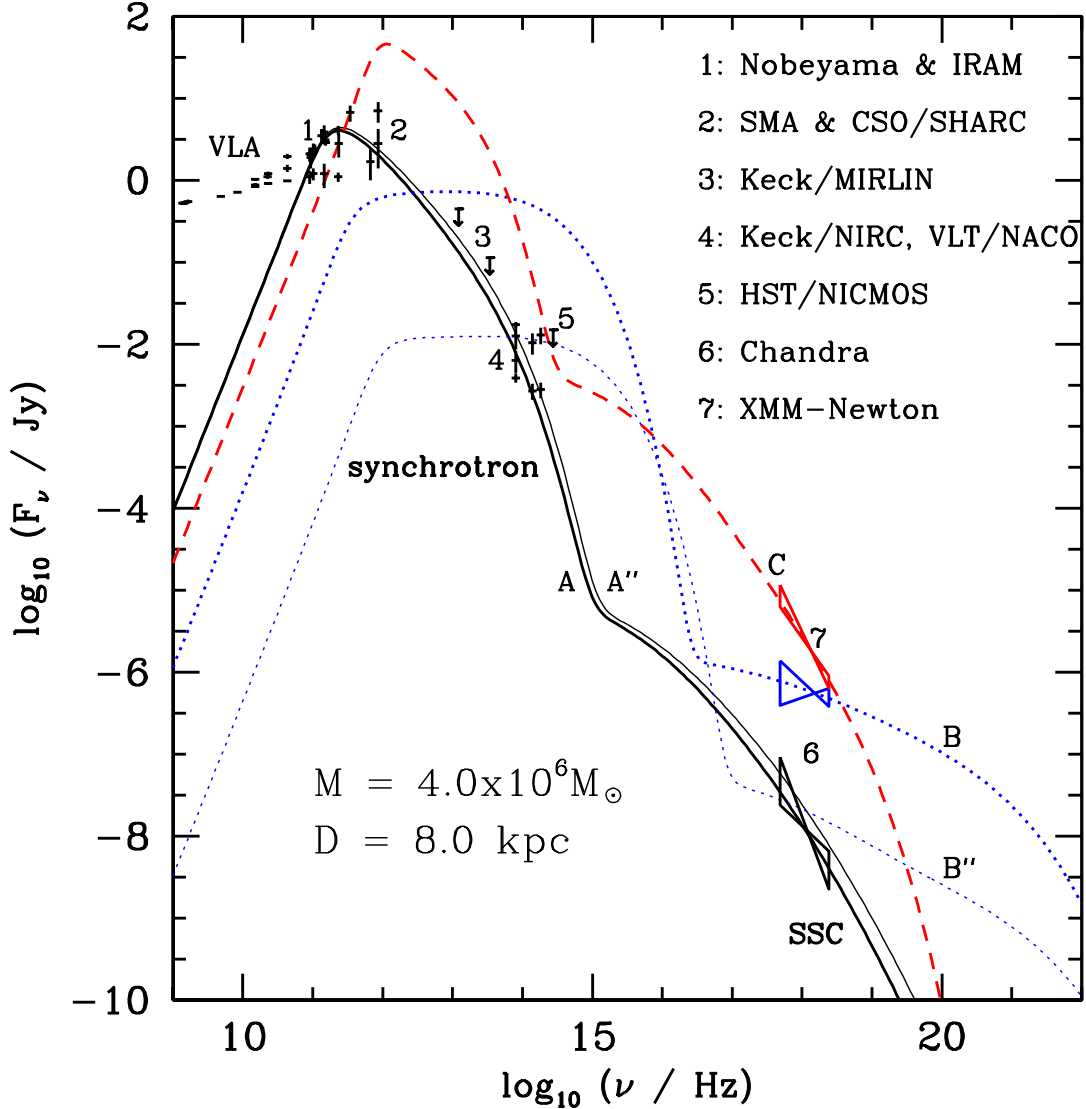}
\caption{
Model fits to the broadband spectrum of Sagittarius A*. The data are gathered from 
observations made at different epochs (the instruments are indicated in the figure by the 
numbers). In the radio and infrared bands, the upper set of data show Sagittarius A*'s 
peak flux densities during the brightest flares observed so far, while at the same
frequencies, the lower data points correspond to its emission in the quiescent state. 
The middle set of data in the radio are the averaged flux densities. The upper butterfly 
in the X-ray band gives the peak flux density for the flare observed by {\it XMM-Newton} 
on October 3 2002. The middle one gives the peak flux density for the 10-27-2000 flare 
observed by {\it Chandra}. The lower one corresponds to the averaged X-ray emission in 
the quiescent state. The upper limits are also for the quiescent state emission. The
spectrum produced by each of the five models in Table \ref{tab:mods} is indicated 
by a continuous curve. See text and the caption of Table \ref{tab:mods} for details 
of the models.\label{fig:spec}
}
\end{figure}

In Table \ref{tab:mods}, we summarize parameters of several models that account for
Sagittarius A*'s spectrum in its various states (Figure \ref{fig:spec}). The corresponding
time scales and normalized electron distributions in the steady state are shown in Figure
\ref{fig:acc} for Models A (left panel), B (middle panel) and C (right panel). The valleys
in the acceleration and the peaks in the escape time correspond to the spectral break 
$k_{\rm max}$ of the turbulence. The loss time $\tau_{\rm loss}\equiv E/\dot{E}_{\rm L}$ 
peaks at an energy, which only depends on the plasma parameter $\alpha$ (see eq. 
[\ref{loss}]). In the lower panels of the figure, we also show the source {\it thermal} 
electron distribution and the flux $f = N/T_{\rm esc}$ of escaping electrons.
It is crucial to note that the acceleration time for Model B (Model C) is shorter than the
duration of X-ray flares with a hard (soft) spectrum. We can therefore use the steady state
electron distribution to calculate the synchrotron and synchrotron self-Comptonization 
(SSC) photon spectra of the models. These are compared with Sagittarius A*'s spectra in 
its various states in Figure \ref{fig:spec}, where the data are gathered from observations 
made at different epochs \citep{Falcke98, Zhao03, Miyazaki99, Miyazaki03, Serabyn97, Dowell03, 
Cotera99, Stolovy03, Baganoff031}.

{\bf Model A} explains Sagittarius A*'s emission at frequencies above $100$ GHz in the
quiescent state. The mm to infrared emission is produced via synchrotron processes,
whereas the optical to gamma-ray emission is produced by SSC (see Melia et al. 2000).  
Global fluctuations in the PWT will vary $\tau_{\rm p}$ and result in a different spectrum. 
For example, by increasing $\tau_{\rm p}^{-1}$ by $5\%$, we obtain {\bf Model 
A$^{\prime\prime}$}, which predicts a weak flare with a soft 
spectrum. The flux densities in the infrared and X-ray bands increase by $\sim80\%$.  
Such a flare is clearly detectable in the infrared by VLT and Keck, but not by the 
existing X-ray instruments due to its low flux. If the flare only lasted for a few hours, 
neither {\it Chandra} nor {\it XMM-Newton} would be able to detect it. Actually, 
Sagittarius A*'s quiescent state X-ray spectrum is obtained by averaging over a long 
observation period \citep{Baganoff031}, which will smooth out contribution from such weak 
flares\footnote{Just before the submission of this paper, we learnt that coordinated IR and 
X-ray observations seem to have detected a similar flare (GCNEWS Vol.17; 
http://www.aoc.nrao.edu/~gcnews)}. {\it This aspect of the model therefore may explain why 
the occurrence frequency of infrared flares is more than twice that of X-ray flares} 
\citep{Genzel03}.

{\bf Model B (B$^{\prime\prime}$)} reproduces X-ray (and probably infrared) flares with
hard spectra. Compared with Model A (A$^{\prime\prime}$), it has a much smaller radius,
which suggests that this is a localized (as opposed to global) event. The quiescent state
emission of Model A presumably still exists during this type of flare. If the plasma is in
a Keplerian motion (see Melia et al. 2001), one would expect to detect quasi-periodic
variability, as has been seen in the infrared \citep{Genzel03}, and apparently now in
X-rays as well (Aschenbach et al. 2004). Note that in the energy range
$80\lesssim\gamma\lesssim2000$ the acceleration time is shorter than the escape time
(middle panel of Figure \ref{fig:acc}). This means that the acceleration of electrons in
this range is more efficient than at lower energies, resulting in a harder electron
distribution and corresponding harder photon spectrum. These spectra are cut off sharply
where $\tau_{\rm loss}, T_{\rm esc}<\tau_{\rm a}$. Such flares are likely produced by a
local fluctuation (possibly a reconfiguration or reconnection of the magnetic field), which
produces strong PWT, trapping electrons within the acceleration site more efficiently. This 
model explains the 10-27-2000 flare observed by {\it Chandra} and predicts strong infrared 
emission accompanying the X-ray flare. Model B$^{\prime\prime}$ is obtained by decreasing 
$R$ by a factor of $13$ while keeping $n R\propto (\tau_{\rm a}T_{\rm esc})^{1/2}/\tau_{\rm 
loss}$ and $R/\tau_{\rm p}\propto (T_{\rm esc}/\tau_{\rm a})^{1/2}$ 
the same as in Model B, so that the spectrum does not change significantly. This model 
predicts weaker infrared and X-ray flares, but the spectra are harder, which may explain 
the weak X-ray flares observed by {\it Chandra} \citep{Baganoff032}, and the recently 
observed infrared flares \citep{Genzel03, Ghez03}.

In {\bf Model C}, which explains the 10-3-1002 {\it XMM-Newton} flare, the synchrotron 
loss dominates at very low energy ($\gamma\sim 200$). The corresponding electron 
distribution and X-ray spectrum are steeper. Given the required high density and large 
volume for this model, this type of event is likely induced by an enhancement of the 
accretion rate onto the black hole. Future multi-wavelength observations will be able to 
test the high flux of sub-mm and IR radiation predicted by this model.

\section{Summary and Discussions}
\label{dis}

Based on the theory of stochastic particle acceleration, we have built a model to account 
for Sagittarius A*'s emission at mm and shorter wavelengths. The quiescent state emission 
is attributed to electrons accelerated by turbulence in a magnetized accretion torus 
and flares can be produced via two distinct mechanisms. The IR and X-ray flares with harder 
spectrum are likely induced by a local MHD process, such as magnetic reconnection. Global 
fluctuations generally produce flares with IR and X-ray spectral indexes close to their 
corresponding value in the quiescent state. The model not only accounts for the 
varied spectra of the flares, but also explains their relative occurrence rate observed at 
IR and X-ray.

Radio emission at longer wavelengths cannot be produced within such a small emission 
region \citep{Liu01a}. This is not surprising, given that the radio emission is variable
on a longer time scale of tens of hours to one week, suggesting that this radiation is 
produced at larger radii than what we have been considering here \citep{Zhao04, Zhao03, 
Zhao93, Bower02}. Very interestingly, our model suggests a significant outflow of 
high-energy electrons \citep{Liu02b}, which may very well be the particles that eventually 
produce Sagittarius A*'s cm spectrum via synchrotron emission on a spatially larger scale. 
In the quiescent state (Model A), the power carried away by electrons with $\gamma>100$
(electrons with lower energy may be trapped by the gravitational potential of the
black hole and will not diffuse to larger radii) is about $2\times10^{37}$ ergs 
s$^{-1}$, which is more than enough to power the observed radio emission, whose 
luminosity is about $10^{34-35}$ ergs s$^{-1}$. (Note that in this model, the mass 
accretion rate is $\sim 10^{18-19}$g s$^{-1}$, for which a few percent of the dissipated 
gravitational energy is carried away by the outwardly diffusing electrons; see the caption 
of Table \ref{tab:mods}.) For the {\it XMM-Newton} flare on October 3 2002 (Model C), the 
total energy carried away by the escaping high-energy electrons is about $7\times 10^{40}$ 
ergs which for a few days can sustain the radio flare observed 13 hours after the X-ray 
event \citep{Zhao04}. This picture also explains the weak correlation between the {\it 
Chandra} X-ray flares and the radio emission from Sagittarius A* \citep{Baganoff032}, 
because the flux of high energy electrons produced during these flares is much smaller than 
that for the {\it XMM-Newton} flare;  no strong enhancement of radio emission is expected 
following a {\it Chandra}-type of X-ray flare. 

One of the intriguing properties of Sagittarius A*'s radio emission is that it is
circularly polarized below $100$ GHz, even though no linear polarization has been observed
there \citep{Bower02}. However, in the mm and sub-mm range only strong linear polarization
is observed \citep{Aitken00, Bower03}. The mm/sub-mm polarization is likely associated with
the structure of the magnetic field at small radii \citep{Agol00, Melia00, Bromley01},
which is consistent with the spectral formation at these wavelengths discussed in this
paper. The observed circular polarization may be due to anisotropy of the escaping 
electrons as they are transported along magnetic field lines toward larger radii under
the influence of synchrotron losses \citep{McTiernan90}. Relativistic electrons beamed
along magnetic field lines produces synchrotron emission with significant circular
polarization while the degree of linear polarization can be suppressed by irregularities
in the source magnetic field \citep{Epstein73a, Epstein73b}. Work to demonstrate this 
feature self-consistently is in progress, and the results will be reported elsewhere. The 
model we have presented here shows promise in being able to account not only for the 
spectral characteristics of Sagittarius A*, but also for its polarization characteristics
and time variation properties.

Finally, we emphasize that although we suggested that $\eta$, $k_{\rm max}/\Omega_{\rm e}$, 
$\alpha$, and $a$ may not change significantly over time, and have fixed them in all the 
models discussed above, given the dramatic changes of the physical conditions during a 
flare state, these parameters may also vary slightly, giving rise to different emission 
spectra than those presented here. Indeed, if the infrared upper limits reported by 
Hornstein et al. (2002) are confirmed, one then needs to decrease the high frequency 
cutoff of the synchrotron emission by adjusting these parameters. This will be 
investigated in a more comprehensive study of particle acceleration in Sagittarius A*.

\acknowledgments

SL thanks Jun-Hui Zhao for helpful discussion and providing the sub-mm data and Macro 
Fatuzzo for his assistance on the Comptonization code. This research was partially 
supported by NSF grant ATM-0312344, NASA grants NAG5-12111, and NAG5 11918-1 (at Stanford), 
and NASA grants NAG5-8239, NAG5-9205, and NAG5-8277 (at Arizona). FM is very grateful to 
the University of Melbourne for its support (through a Miegunyah Fellowship).

\clearpage

\clearpage



\begin{table}
\begin{center}
\caption{Models and Parameters.\label{tab:mods}}
\vskip 0.2in
\begin{tabular}{l|ccc|ccc}
\tableline\tableline
Models   & $R$($r_S$) & $\tau_{\rm p}^{-1}$ (s$^{-1}$) & $n$ 
($10^7$cm$^{-3}$) & 
$B$ (Gauss)& $k_{\rm b}T$ ($m_{\rm e}c^2$)  & $Q$ ($10^{42}$s$^{-1}$) \\
\tableline
A (A$^{\prime\prime}$)    & 2.5        & 0.74 (0.78) & 0.76 & 8.8 &  26.0 
(27.3) & 3.2 (3.0) 
\\
B (B$^{\prime\prime}$)    & 0.22 (0.017)\tablenotemark{\dag}& 26.7 (345)  & 19.0 (240) 
& 44 
(160) & 34.2 (20.5) & 
0.11 (0.014) \\
C        & 1.9        & 1.78       & 18.0      & 43      & 10.5 & 49\\
\tableline
\end{tabular}
\tablenotetext{\dag}{The radius $R$ is the radius of the assumed spherical
emitting region. In some models, $R<r_S$, meaning that the fluctuation producing
the flare is not global, but rather is localized near the black hole.  See text
for details.}



\tablecomments{A sampling of models for Sagittarius A* and their corresponding 
parameters. Note that the first three parameters are the primary parameters. All other 
quantities are derived from them. $Q$ is the steady state injection rate. For Models 
A (A$^{\prime\prime}$) and C, the mass accretion rate onto the black hole can be estimated 
as ${\dot M}\simeq Qm_{\rm p}$ because high energy electrons ($\gamma>100$), which may 
diffuse to larger radii, account for less than $10\%$ of $Q$. See text for details. Models 
A, B and C correspond to the left, middle and right panels of Figure \ref{fig:acc} and the 
thick solid, dotted and dashed curves in Figure \ref{fig:spec}, which fit Sagittarius 
A*'s emission in the quiescent state, at the peak of the 10-27-2000 {\it Chandra} flare 
and the 10-3-2002 {\it XMM-Newton} flare, respectively. The parameters for model 
A$^{\prime\prime}$ and B$^{\prime\prime}$ are indicated in the parentheses 
whenever different. The 
corresponding emission spectra are indicated, respectively, by the thin solid and 
dotted curves in Figure \ref{fig:spec}, which explain the weaker infrared and X-ray 
flares.
}
\end{center}
\end{table}






\end{document}